\documentclass[prb]{revtex4}

\begin{document}
\title{Comment on ``Multi-phonon Raman scattering in semiconductor
nanocrystals: Importance of non-adiabatic transitions''}
\author{E. \surname{Men\'{e}ndez-Proupin}}
\email[Email: ]{eariel@ff.oc.uh.cu}
\affiliation{Institute of Materials and Reagents, University of Havana, San Lazaro y L, Vedado 10400, Havana, Cuba}
\date{\today}
\begin{abstract}
In a recent paper [E. P. Pokatilov et al, Phys. Rev. B \textbf{65}, 075316 (2002)] the 
Raman selection rules in spherical nanocrystals with a degenerate valence band are analyzed. Some precisions are given here.
\end{abstract}
\maketitle

The authors state that the one-phonon Raman spectrum of spherical CdSe
nanocrystals in crossed polarization configuration is determined only by
d-phonons with angular momentum projection $m=\pm 1$. This is not a general 
result, as the number $m$
is rather conventional. First, to define the angular momentum projection, we
need to set a quantization axis, respect to which the phonons are defined.
Let the quantization axis be the $Z$-axis of our coordinate system. If we
choose another quantization axis, let say $Z^{\prime }$, we shall find new $%
m^{\prime }=\pm 1$ phonons, which are linear combinations of $Z$-quantized
phonons with all values of $m$.

Now, consider a Raman experiment with incident an scattered light linearly
and crossly polarized. We may choose the $Z$- and $Y$-axis of our reference
frame to coincide with the incident and scattered polarization vectors,
respectively. According to the theory of angular momentum and the dipole
approximation, the absorption of an incident $Z$-polarized photon creates a $%
P$-type electronic excitation with angular momentum projection $M_{1}=0$.
The nature of the electronic excitation is not important for this analysis.
Furthermore, the emission of a $Y$-polarized photon is possible only by
annihilation of a $P$-type electronic excitation with $M_{2}=\pm 1$. The
transition between these electronic excitations in the case of CdSe
nanocrystals is carried out through the creation or annihilation of phonons
with $m=\pm 1$, as stated by the authors. Now, let rotate the reference 
frame, so that the new $%
X^{\prime }$- and $Y^{\prime }$-axis of coincide with the incident and
scattered polarization vectors, respectively. 
Then, the electronic excitations created
when the incident photon is absorbed can have $M_{1}^{\prime }=\pm 1$. The
transitions to $M_{2}^{\prime }=\pm 1$ electronic states can be mediated
only by phonons with $m^{\prime }=0,\pm 2$ ($M_{2}^{\prime }-M_{1}^{\prime
}=0,\pm 2$).\ 

Both descriptions describe the same reality: the calculated Raman cross
section, adding up the contributions of all the phonons, has the same value
in both cases. The distinction between phonons with different angular
projections is purely conventional if the system has spherical symmetry.

Due to the same reasons, in parallel polarization configuration, the Raman
spectrum has contributions from phonons with $m\neq 0$ if the quantization
axis does not coincide with the light polarization vectors.

\end{document}